\documentstyle[12pt,amssymb,amsmath]{article}
\newtheorem{theorem}{Theorem}[section]
\newtheorem{lemma}{Lemma}[section]

\begin{document}

\title{ {\bf Generalized Darboux transformations for the KP equation
with self-consistent sources  }  }
\author{ {\bf  Ting Xiao
    \hspace{1cm} Yunbo Zeng\dag } \\
    {\small {\it
    Department of Mathematical Sciences, Tsinghua University,
    Beijing 100084, China}} \\
    {\small {\it \dag
     Email: yzeng@math.tsinghua.edu.cn}}}

\date{}
\maketitle
\renewcommand{\theequation}{\arabic{section}.\arabic{equation}}

\begin{abstract}
The KP equation with self-consistent sources (KPESCS) is treated
in the framework of the constrained KP equation. This offers a
natural way to obtain the Lax representation for the KPESCS. Based
on the conjugate Lax pairs, we construct the generalized binary
Darboux transformation with arbitrary functions in time $t$ for
the KPESCS which, in contrast with the binary Darboux
transformation of the KP equation, provides a
non-auto-B\"{a}cklund transformation between two KPESCSs with
different degrees. The formula for N-times repeated generalized
binary Darboux transformation is proposed and enables us to find
the N-soliton solution and lump solution as well as some other
solutions of the KPESCS.
\end{abstract}

\hskip\parindent


\section{Introduction}
\setcounter{equation}{0} \hskip\parindent
 Soliton equations with self-consistent sources (SESCSs) are important models in many
fields of physics, such as hydrodynamics, solid state physics,
plasma physics, etc. [1-7]. Until now, much development has been
made in the study of SESCS. For example, in (1+1)-dimensional
case, some SESCSs such as the KdV, modified KdV, nonlinear
Schrodinger, AKNS and Kaup-Newell hierarchies with self-consistent
sources were solved by the inverse scattering method [1,2,3,6-10].
Recently, generalized binary Darboux transformations with
arbitrary functions in time $t$ for some (1+1)-dimensional SESCSs
, which offer a non-auto-B\"{a}cklund transformation between two
SESCSs with different degrees of sources, have been constructed
and can be used to obtain N-soliton solution [11-13].
But in (2+1)-dimensional case, less results to the SESCSs have
been obtained. The KP equation with self-consistent sources
(KPESCS) arose in some physical models describing the interaction
of long and short waves and the soliton solution of it was first
found by Mel'nikov \cite{Mel'nikov87,Mel'nikov89(2)}. However,
since the explicit time part of the Lax representation of the
KPESCS was not found, the method to solve the KPESCS by inverse
scattering transformation in \cite{Mel'nikov87,Mel'nikov89(2)} was
quite complicated. Recently, N-soliton solution of the KPESCS was
obtained by Hirota method in \cite{Zhangdajun2003}. In this paper,
we treat the constrained KP hierarchy as the stationary equations
of the KP hierarchy with self-consistent sources. This gives a
natural way to find the Lax pair for the KPESCS. Using the
conjugate Lax pairs, we construct the generalized binary Darboux
transformation with arbitrary functions in time $t$ for the
KPESCS. In contrast with the binary Darboux transformation for the
KP equation which offers
 B\"{a}cklund transformation, this transformation
provides a non-auto-B\"{a}cklund transformation between two
KPESCSs with different degrees of sources. Some interesting
solutions of KPESCS such as soliton solution, lump solution and
mixture solution of exponential and rational solutions are
obtained by this generalized Darboux transformation.

  The paper will be organized as follows. We recall some facts about the
Darboux transformation for the KP equation in the next section. In
section 3, through the pseudo-differential operator (PDO)
formalism we reveal the relation between the KP hierarchy with
self-consistent sources and the constrained KP hierarchy. Then the
conjugate Lax pairs of the KP hierarchy with self-consistent
sources can be obtained naturally. Using the conjugate Lax pairs,
we can construct the generalized Darboux transformations with
arbitrary functions in time for KPESCS and find some interesting
solutions of the KPESCS. In Section 4, the N-times repeated
generalized Darboux transformation will be constructed by which
the N-soliton solution and some other solutions can be obtained.

\section{The Darboux transformation for the KP equation}
\setcounter{equation}{0} \hskip\parindent We first give a simple
description of the KP hierarchy within the framework of the Sato
theory (see \cite{Date83,Ohta1988}). Let us consider the following
pseudo-differential operator(PDO)
\begin{equation}
\label{a2}
    L=\partial+u_0\partial^{-1}+u_1\partial^{-2}+...,
\end{equation}
where $\partial$ denotes $\frac{\partial}{\partial x}$, and
$u_j,j=0,1,...$ are functions. Denote $B_m=(L^m)_+$ for $\forall m
\in N$ where $(L^m)_+$ represents the projection of $L^m$ to its
pure differential part. Then the KP hierarchy has the following
Lax representation (zero-curvature representation)
\begin{equation}
\label{a3}
    (B_n)_{t_k}-(B_k)_{t_n}+[B_n,B_k] = 0,\ \ n,k\geq2.
\end{equation}
The equation (\ref{a3}) has a pair of conjugate Lax pairs as
follows
\begin{subequations}
\label{a4}
\begin{equation}
\label{a41}
    \psi^-_{t_k} = B_k\psi^-,
\end{equation}
\begin{equation}
\label{a42}
    \psi^-_{t_n} = B_n\psi^-,
\end{equation}
\end{subequations}
and
\begin{subequations}
\label{a5}
\begin{equation}
\label{a51}
    \psi^+_{t_k} = -(B_k)^*\psi^+,
\end{equation}
\begin{equation}
\label{a52}
    \psi^+_{t_n} = -(B_n)^*\psi^+.
\end{equation}
\end{subequations}

When $k=2,\ n=3$ and under the following transformation
\begin{equation}
\label{a7}
    u=2u_0,\ \  t=-\frac{1}{4}t_3,\ \  y=\alpha t_2,
\end{equation}
we get the simplest and most important equation in the hierarchy
(\ref{a3}), the KP equation
\begin{equation}
\label{a1}
    (u_t+6uu_x+u_{xxx})_x+3\alpha^2u_{yy} =0,
\end{equation}
which appears in physical application in two different forms with
$\alpha=1$ and $\alpha=i$, usually referred to as the KPI and KPII
equation \cite{Ablowitz91}. Under the transformation (\ref{a7}),
we obtain the conjugate Lax pairs of (\ref{a1}) respectively from
(\ref{a5}) and (\ref{a4}) as follows
\begin{subequations}
\label{eqn:a2}
\begin{equation}
\label{eqn:a21}
    \alpha\psi^+_y=-\psi^+_{xx}-u\psi^+,
\end{equation}
\begin{equation}
\label{eqn:a22}
    \psi^+_t=A^+(u)\psi^+,\ \ \
    A^+(u)=-4\partial^3-6u\partial-3(u_x-\alpha\partial^{-1}u_y),
\end{equation}
\end{subequations}
and
\begin{subequations}
\label{eqn:a3}
\begin{equation}
\label{eqn:a31}
    \alpha\psi^-_y=\psi^-_{xx}+u\psi^-,
\end{equation}
\begin{equation}
\label{eqn:a32}
    \psi^-_t=A^-(u)\psi^-,\ \  A^-(u)=-4\partial^3-6u\partial-3(u_x+\alpha\partial^{-1}u_y).
\end{equation}
\end{subequations}

From (\ref{eqn:a2}) and (\ref{eqn:a3}), we can construct three
types of Darboux transformations for the KP equation (\ref{a1}).
\\
(1)The forward Darboux transformation for the KP equation.

Assume that $u$ be a solution of the KP equation (\ref{a1}) and
denote a fixed solution of (\ref{eqn:a2}) by
$\psi^+_1=\psi^+_1(x,y,t)$. The forward Darboux transformation for
(\ref{eqn:a2}) is given by \cite{Matveev91}
\begin{subequations}
\label{eqn:a4}
\begin{equation}
\label{eqn:a41}
    \psi^+[+1]=\psi^+_{x}-\frac{\psi^+_{1,x}}{\psi^+_1}\psi^+,
\end{equation}
\begin{equation}
\label{eqn:a42}
    u[+1]=u+2\partial^2\mathrm{ln}\psi^+_1.
\end{equation}
\end{subequations}
So $u[+1]$ is a new solution of the KP equation (\ref{a1}).
Substituting (\ref{eqn:a4}) into (\ref{eqn:a22}), we have
\begin{equation}
\label{a6}
\begin{array}{lll}
    A^+(u[+1])\psi^+[+1]&=&[\psi^+_{x}-\frac{\psi^+_{1,x}}{\psi^+_1}\psi^+]_t
    \\
    &=&\psi^+_{xt}-(\frac{\psi^+_{1,x}}{\psi^+_1}\psi^+)_t
    \\
    &=&(A^+(u)\psi^+)_x-(\frac{A^+(u)\psi^+_1}{\psi^+_1})_x\psi^+
    -(\frac{\psi^+_{1,x}}{\psi^+_1})A^+(u)\psi^+ .
\end{array}
\end{equation}
(2)The backward Darboux transformation for the KP equation.

Assume that $u$ be a solution of the KP equation (\ref{a1}) and
denote a fixed solution of (\ref{eqn:a3}) by
$\psi^-_2=\psi^-_2(x,y,t)$. The backward Darboux transformation
for the system (\ref{eqn:a2}) is defined by
\begin{subequations}
\label{eqn:a7}
\begin{equation}
\label{eqn:a71}
    \psi^+[-1]=\frac{a_1+\int\psi^+\psi^-_2\mathrm{d}x}{\psi^-_2},
\end{equation}
\begin{equation}
\label{eqn:a72}
    u[-1]=u+2\partial^2\mathrm{ln}\psi^-_2,
\end{equation}
\end{subequations}
where $a_1$ is an arbitrary constant. We point out that throughout
the paper, the integral operation $\int f_1f_2 \mathrm{d}x$ such
as $\int\psi^+\psi^-_2\mathrm{d}x$ here means $\int_{-\infty}^x
f_1f_2 \mathrm{d}x$ or $-\int_x^{\infty} f_1f_2 \mathrm{d}x$ and
contains no arbitrary function of $y$ and $t$, only numerical
constant if we impose some suitable boundary condition on the
integrand functions $f_1$ and $f_2$ at $x=-\infty$ or $x=\infty$.
For arbitrariness of the constants in the Darboux transformations
such as $a_1$ here, in our computation later, the integral
constants are taken to be zero.

Substituting (\ref{eqn:a7}) into (\ref{eqn:a22}), we get the
following equality
\begin{equation}
\label{a9}
\begin{array}{lll}
    A^+(u[-1])\psi^+[-1]&=&(\frac{a_1+\int\psi^+\psi^-_2\mathrm{d}x}{\psi^-_2})_t
    \\
    &=&-\frac{A^-(u)\psi^-_2}{(\psi^-_2)^2}(a_1+\int\psi^+\psi^-_2\mathrm{d}x)
    +\frac{\int[\psi^+(A^-(u)\psi^-_2)+\psi^-_2(A^+(u)\psi^+)]\mathrm{d}x}{\psi^-_2}.
\end{array}
\end{equation}
(3)The binary Darboux transformation for the KP equation.

 When the backward Darboux transformation and forward
 Darboux transformation are applied consecutively to the system
 (\ref{eqn:a2}), we can get the binary DT as follows
\begin{subequations}
\label{eqn:a10}
\begin{equation}
\label{eqn:a101}
     \psi^+[-1,+1]=\psi^+[-1]_{x}-\frac{\psi^+_1[-1]_x}{\psi^+_1[-1]}\psi^+[-1]=\psi^+-\frac{\psi^+_1(a_1+\int\psi^+\psi^-_2\mathrm{d}x)}{a_2+\int\psi^-_2\psi^+_1\mathrm{d}x},
\end{equation}
\begin{equation}
\label{eqn:a102}
    u[-1,+1]=u[-1]+2\partial^2\mathrm{ln}\psi^+_1[-1]=u+2\partial^2\mathrm{ln}(a_2+\int\psi^-_2\psi^+_1\mathrm{d}x),
\end{equation}
\end{subequations}
where
$\psi^+_1[-1]=\frac{a_2+\int\psi^+_1\psi^-_2\mathrm{d}x}{\psi^-_2}$,\
\ $a_j$,\ \ $j=1,2$ are arbitrary constants.
\\\\
{\bf{Remark}}: To system (\ref{eqn:a3}), we can define the forward
DT, backward DT and binary DT for it analogously. We list the
results here:
\\
Forward DT of (\ref{eqn:a3}),
\begin{subequations}
\label{eqn:a11}
\begin{equation}
\label{eqn:a111}
     \psi^-\rightarrow　\psi^-[-1]=\psi^-_x-\frac{\psi^-_{2,x}}{\psi^-_2}\psi^-,
\end{equation}
\begin{equation}
\label{eqn:a112}
    u\rightarrow　u[-1]=u+2\partial^2\mathrm{ln}\psi^-_2,
\end{equation}
\end{subequations}
backward DT of (\ref{eqn:a3}),
\begin{subequations}
\label{eqn:a12}
\begin{equation}
\label{eqn:a121}
     \psi^-\rightarrow　\psi^-[+1]=\frac{a_3+\int\psi^-\psi^+_1\mathrm{d}x}{\psi^+_1},
\end{equation}
\begin{equation}
\label{eqn:a122}
    u\rightarrow　u[+1]=u+2\partial^2\mathrm{ln}\psi^+_1,
\end{equation}
\end{subequations}
Binary DT of (\ref{eqn:a3}),
\begin{subequations}
\label{eqn:a13}
\begin{equation}
\label{eqn:a131}
     \psi^-\rightarrow　\psi^-[+1,-1]=\psi^--\frac{\psi^+_1(a_3+\int\psi^-\psi^+_1\mathrm{d}x)}{a_2+\int\psi^-_2\psi^+_1\mathrm{d}x},
\end{equation}
\begin{equation}
\label{eqn:a132}
    u\rightarrow　u[+1,-1]=u+2\partial^2\mathrm{ln}(a_2+\int\psi^+_1\psi^-_2\mathrm{d}x),
\end{equation}
\end{subequations}
where
$\psi^-_2[+1]=\frac{a_2+\int\psi^+_1\psi^-_2\mathrm{d}x}{\psi^+_1}$,\
\ $a_j$,\ $j=2,3$ are arbitrary constants.

\section{The Darboux transformation for the KP equation with self-consistent sources}
\setcounter{equation}{0} \hskip\parindent
 We first recall the constrained KP
 hierarchy (more details can be referred to [22-26] and the references
 therein). For the pseudo-differential operator $L$ given by (\ref{a2}), if we impose
the condition that $(q_j)_{t_k}=B_kq_j$,\
$(r_j)_{t_k}=-B_k^*q_j$,\ $j=1,...,m, k\geq 2$, and make the
constraint
$$(L^n)_-=\sum_{j=1}^mq_j\partial^{-1}r_j,\ \ or\ \  L^n=B_n+\sum_{j=1}^mq_j\partial^{-1}r_j,$$
the n-constrained KP hierarchy is defined as follows
\begin{subequations}
\label{b01}
\begin{equation}
\label{b011}
     (L^n)_{t_k}=[B_k,L^n],
\end{equation}
\begin{equation}
\label{b012}
    (q_j)_{t_k}=B_kq_j,
\end{equation}
\begin{equation}
\label{b013}
    (r_j)_{t_k}=-B_k^*r_j,\ \ \ j=1,...,m.
\end{equation}
\end{subequations}

If the term $(B_k)_{t_n}$ is added to the right hand side of
equation (\ref{b011}), we get the KP hierarchy with
self-consistent sources as
\begin{subequations}
\label{b02}
\begin{equation}
\label{b021}
     (B_k)_{t_n}-(L^n)_{t_k}+[B_k,L^n]=0,
\end{equation}
\begin{equation}
\label{b022}
    (q_j)_{t_k}=B_kq_j,
\end{equation}
\begin{equation}
\label{b023}
    (r_j)_{t_k}=-B_k^*r_j,\ \ \ j=1,...,m.
\end{equation}
\end{subequations}

If the variable "$t_n$" is viewed as the evolution variable, the
n-constrained KP hierarchy (\ref{b01}) may be considered as the
stationary case (i.e. $(B_k)_{t_n}=0$) of the KP hierarchy with
self-consistent sources (\ref{b02}). Under the condition
(\ref{b022}) and (\ref{b023}), we naturally get the conjugate Lax
pairs for (\ref{b021}) as
\begin{subequations}
\label{b03}
\begin{equation}
\label{b031}
     \psi^-_{t_k} = B_k\psi^-,
\end{equation}
\begin{equation}
\label{b032}
    \psi^-_{t_n} = L^n\psi^- = B_n\psi^-+\sum_{j=1}^mq_j\int r_j\psi^-\mathrm{d}x,
\end{equation}
\end{subequations}
and
\begin{subequations}
\label{b04}
\begin{equation}
\label{b041}
     \psi^+_{t_k} = -B^*_k\psi^+,
\end{equation}
\begin{equation}
\label{b042}
    \psi^+_{t_n}=-(L^n)^*\psi^+ =
    -B_n^*\psi^++\sum_{j=1}^mr_j\int
    q_j\psi^+\mathrm{d}x.
\end{equation}
\end{subequations}

When $k=2,n=3$, under the transformation (\ref{a7}) and
$\Psi_j=-r_j$,\ $\Phi_j=q_j$, we will get the KPESCS\
\cite{Mel'nikov87} from (\ref{b02}) as follows
\begin{subequations}
\label{b1}
\begin{equation}
\label{b11}
     [u_t+6uu_x+u_{xxx}+8(\sum_{j=1}^{m}\Phi_j\Psi_j)_x]_x+3\alpha^2u_{yy} =0,
\end{equation}
\begin{equation}
\label{b12}
    \alpha\Phi_{j,y}=\Phi_{j,xx}+u\Phi_j,
\end{equation}
\begin{equation}
\label{b13}
    \alpha\Psi_{j,y}=-\Psi_{j,xx}-u\Psi_j, \ \ \ j=1,...,m,
\end{equation}
\end{subequations}
and its conjugate Lax pairs from (\ref{b04}) and (\ref{b03})
respectively as follows
\begin{subequations}
\label{b2}
\begin{equation}
\label{b21}
    \alpha\psi^+_y=-\psi^+_{xx}-u\psi^+,
\end{equation}
\begin{equation}
\label{b22}
    \psi^+_t=A^+(u)\psi^++T^+_m(\Psi,\Phi)\psi^+,\
    T^+_m(\Psi,\Phi)\psi^+=4\sum_{j=1}^{m}\Psi_j\int\Phi_j\psi^+\mathrm{d}x,
\end{equation}
\end{subequations}
and
\begin{subequations}
\label{b3}
\begin{equation}
\label{b31}
    \alpha\psi^-_y=\psi^-_{xx}+u\psi^-,
\end{equation}
\begin{equation}
\label{b32}
    \psi^-_t=A^-(u)\psi^-+T^-_m(\Psi,\Phi)\psi^-,\ T^-_m(\Psi,\Phi)\psi^-=4\sum_{j=1}^{m}\Phi_j\int\Psi_j\psi^-\mathrm{d}x.
\end{equation}
\end{subequations}

From the conjugate Lax pairs for KPESCS (\ref{b1}), we can also
construct the forward DT, backward DT and binary DT for it.
\begin{theorem}
 Assume $u, \Phi_1,...,\Phi_m,\Psi_1,...,\Psi_m$ be the solution of the KPESCS (\ref{b1})
  and $\psi^+_1$ satisfies (\ref{b2}), then the forward Darboux  transformation
   for (\ref{b2}) can be defined by
\begin{subequations}
\label{b4}
\begin{equation}
\label{b41}
\psi^+[+1]=\psi^+_{x}-\frac{\psi^+_{1,x}}{\psi^+_1}\psi^+,
\end{equation}
\begin{equation}
\label{b42} u[+1]=u+2\partial^2\mathrm{ln}\psi^+_1,
\end{equation}
\begin{equation}
\label{b43}
 \Phi_j[+1]=-\frac{\int\psi^+_1\Phi_j\mathrm{d}x}{\psi^+_1},
\end{equation}
\begin{equation}
\label{b44}
 \Psi_j[+1]=\Psi_{j,x}-\frac{\psi^+_{1,x}}{\psi^+_1}\Psi_j,\qquad
 j=1,...,m.
\end{equation}
\end{subequations}
Namely, $\ u[+1],\ \psi^+[+1],\ \Psi_j[+1],\ \Phi_j[+1],\
j=1,...,m,$ satisfy (\ref{b1}) and (\ref{b2}).
\end{theorem}
{\bf{Proof}}:\ Based on the results in the previous section, it is
obvious that (\ref{b12}), (\ref{b13}) and (\ref{b21}) hold under
the transformation (\ref{b4}). So we only need to prove
(\ref{b22}), i.e., the following equality,
\begin{eqnarray}
\label{b7} \psi^+[+1]_t
&=&[\psi^+_{x}-\frac{\psi^+_{1,x}}{\psi^+_1}\psi^+]_t
\nonumber\\
&=&(A^+(u)\psi^++T^+_m(\Psi,\Phi)\psi^+)_x-(\frac{A^+(u)\psi^+_1+T^+_m(\Psi,\Phi)\psi^+_1}{\psi^+_1})_x\psi^+
\nonumber\\
& &
-\frac{\psi^+_{1,x}}{\psi^+_1}(A^+(u)\psi^++T^+_m(\Psi,\Phi)\psi^+)
\nonumber\\
&=&(A^+(u)\psi^+)_x-(\frac{A^+(u)\psi^+_1}{\psi^+_1})_x\psi^+-\frac{\psi^+_{1,x}}{\psi^+_1}A^+(u)\psi^+
\nonumber\\
& &
+(T^+_m(\Psi,\Phi)\psi^+)_x-(\frac{T^+_m(\Psi,\Phi)\psi^+_1}{\psi^+_1})_x\psi^+-\frac{\psi^+_{1,x}}{\psi^+_1}T^+_m(\Psi,\Phi)\psi^+
\nonumber\\
&=&A^+(u[+1])\psi^+[+1]+T^+_m(\Psi[+1],\Phi[+1])\psi^+[+1].
\end{eqnarray}

By simple computation, we can prove that equality (\ref{a6}) still
holds now, so we only need to check the terms containing
$\Phi_j,\Psi_j$ in the equality (\ref{b7}), i.e., to check the
following equality
\begin{equation}
\label{b8}
(T^+_m(\Psi,\Phi)\psi^+)_x-(\frac{T^+_m(\Psi,\Phi)\psi^+_1}{\psi^+_1})_x\psi^+-\frac{\psi^+_{1,x}}{\psi^+_1}T^+_m(\Psi,\Phi)\psi^+=T^+_m(\Psi[+1],\Phi[+1])\psi^+[+1].
\end{equation}

In fact, we have
\begin{equation}
\label{b9}
\begin{array}{ll}
&\text{the l.h.s. of (\ref{b8})}
\\
=&4[\sum_{j=1}^{m}\Psi_j\int\Phi_j\psi^+\mathrm{d}x]_x-4[\frac{\sum_{j=1}^{m}\Psi_j\int\Phi_j\psi^+_1\mathrm{d}x}{\psi^+_1}]_x\psi^+-4\frac{\psi^+_{1,x}}{\psi^+_1}\sum_{j=1}^{m}\Psi_j\int\Phi_j\psi^+\mathrm{d}x
\\
=&4\sum_{j=1}^{m}[\Psi_{j,x}\int\Phi_j\psi^+\mathrm{d}x+\Psi_j\Phi_j\psi^+]-4\frac{\psi^+_{1,x}}{\psi^+_1}\sum_{j=1}^{m}\Psi_j\int\Phi_j\psi^+\mathrm{d}x
\\
&-4\frac{\sum_{j=1}^{m}[(\Psi_{j,x}\int\Phi_j\psi^+_1\mathrm{d}x+\Psi_j\Phi_j\psi^+_1)\psi^+_1-(\Psi_j\int\Phi_j\psi^+_1\mathrm{d}x)\psi^+_{1,x}]}{(\psi^+_1)^2}\psi^+
\\
=&4\sum_{j=1}^{m}[\Psi_{j,x}\int\Phi_j\psi^+\mathrm{d}x-\frac{\psi^+\Psi_{j,x}}{\psi^+_1}\int\Phi_j\psi^+_1\mathrm{d}x+\Psi_j\psi^+\psi^+_{1,x}\frac{\int\Phi_j\psi^+_1\mathrm{d}x}{(\psi^+_1)^2}-\frac{\psi^+_{1,x}}{\psi^+_1}\Psi_j\int\Phi_j\psi^+\mathrm{d}x]
\\
=&4\sum_{j=1}^{m}[(\Psi_{j,x}-\frac{\psi^+_{1,x}}{\psi^+_1}\Psi_j)\int\Phi_j\psi^+\mathrm{d}x-\frac{\psi^+\int\Phi_j\psi^+_1\mathrm{d}x}{\psi^+_1}(\Psi_{j,x}-\frac{\psi^+_{1,x}}{\psi^+_1}\Psi_j)]
\\
=&4\sum_{j=1}^{m}\Psi_j[+1](\int\Phi_j\psi^+\mathrm{d}x-\frac{\psi^+}{\psi^+_1}\int\Phi_j\psi^+_1\mathrm{d}x)
\end{array}
\end{equation}

\begin{equation}
\label{b10}
\begin{array}{lll}
\text{the r.h.s. of
(\ref{b8})}&=&4\sum_{j=1}^{m}\Psi_j[+1]\int\Phi_j[+1]\psi^+[+1]\mathrm{d}x
\\
&=&4\sum_{j=1}^{m}\Psi_j[+1]\int[-\frac{\int\psi^+_1\Phi_j\mathrm{d}x}{\psi^+_1}(\psi^+_x-\frac{\psi^+_{1,x}}{\psi^+_1}\psi^+)]\mathrm{d}x
\\
&=&-4\sum_{j=1}^{m}\Psi_j[+1]\int[(\int\psi^+_1\Phi_j\mathrm{d}x)d(\frac{\psi^+}{\psi^+_1})]\mathrm{d}x
\\
&=&4\sum_{j=1}^{m}\Psi_j[+1][\int\Phi_j\psi^+\mathrm{d}x-\frac{\psi^+}{\psi^+_1}\int\Phi_j\psi^+_1\mathrm{d}x]
\\
&=&\text{the l.h.s. of (\ref{b8})}
\end{array}
\end{equation}
This completes the proof.
\\
\begin{theorem}
 Assume $u, \Phi_1,...,\Phi_m,\Psi_1,...,\Psi_m$ be a solution of the KPESCS (\ref{b1})
  and $\psi^-_2$ satisfies (\ref{b3}), then the backward
  DT for (\ref{b2}) is defined by
\begin{subequations}
\label{c1}
\begin{equation}
\label{c11}
\psi^+[-1]=\frac{a_4+\int\psi^-_2\psi^+\mathrm{d}x}{\psi^-_2},
\end{equation}
\begin{equation}
\label{c12} u[-1]=u+2\partial^2\mathrm{ln}\psi^-_2,
\end{equation}
\begin{equation}
\label{c13}
 \Phi_j[-1]=\Phi_{j,x}-\frac{\psi^-_{2,x}}{\psi^-_2}\Phi_j,
\end{equation}
\begin{equation}
\label{c14}
 \Psi_j[-1]=-\frac{\int\psi^-_2\Psi_j\mathrm{d}x}{\psi^-_2},\qquad j=1,...,m,
\end{equation}
\end{subequations}
where $a_4$ is an arbitrary constant, namely $\ u[-1],\
\psi^+[-1],\ \Psi_j[-1],\ \Phi_j[-1],\  j=1,...,m,$ satisfy
(\ref{b1}) and (\ref{b2}).
\end{theorem}
{\bf{Proof}}: It is obvious that (\ref{b12}),(\ref{b13}) and
(\ref{b21}) hold under the transformation (\ref{c1}). So we only
need to prove (\ref{b22}), i.e., to prove the following equality
\begin{equation}
\label{c4}
\begin{array}{ll}
&\psi^+[-1]_t\\
=&[\frac{a_4+\int\psi^-_2\psi^+\mathrm{d}x}{\psi^-_2}]_t
\\
=&-\frac{A^-(u)\psi^-_2}{(\psi^-_2)^2}(a_4+\int\psi^-_2\psi^+\mathrm{d}x)+\frac{\int[(A^-(u)\psi^-_2)\psi^++(A^+(u)\psi^+)\psi^-_2]\mathrm{d}x}{\psi^-_2}
\\
&-\frac{T^-_m(\Phi,\Psi)\psi^-_2}{(\psi^-_2)^2}(a_4+\int\psi^+_2\psi^+\mathrm{d}x)+\frac{\int[(T^-_m(\Phi,\Psi)\psi^-_2)\psi^++(T^+_m(\Psi,\Phi)\psi^+)\psi^-_2]\mathrm{d}x}{\psi^-_2}
\\
=&A^+(u[-1])\psi^+[-1]+T^+_m(\Psi[-1],\Phi[-1])\psi^+[-1].
\end{array}
\end{equation}
Similarly, using equality (\ref{a9}), we only need to check the
terms containing $\Psi_j$,$\Phi_j$, i.e.,
\begin{equation}
\label{c5}
\begin{array}{l}
 -\frac{T^-_m(\Psi,\Phi)\psi^-_2}{(\psi^-_2)^2}(a_4+\int\psi^-_2\psi^+\mathrm{d}x)+\frac{\int[(T^-_m(\Psi,\Phi)\psi^-_2)\psi^++(T^+_m(\Psi,\Phi)\psi^+)\psi^-_2]\mathrm{d}x}{\psi^-_2}
 \\
 =T^+_m(\Psi[-1],\Phi[-1])\psi^+[-1],
 \end{array}
\end{equation}
i.e.,
\begin{equation}
\label{c6}
\begin{array}{l}
 4\sum_{j=1}^{m}\{-\frac{\Phi_j\int\psi^-_2\Psi_j\mathrm{d}x}{(\psi^-_2)^2}(a_4+\int\psi^-_2\psi^+\mathrm{d}x)+\frac{\int[\psi^+\Phi_j\int(\psi^-_2\Psi_j)+\psi^-_2\Psi_j\int(\psi^+\Phi_j)]\mathrm{d}x}{\psi^-_2}\}
 \\
 =4\sum_{j=1}^{m}\Psi_j[-1]\int\Phi_j[-1]\psi^+[-1]\mathrm{d}x.
 \end{array}
\end{equation}

In fact, we have
\begin{eqnarray}
\label{c7} \text{the l.h.s. of
(\ref{c6})}&=&4\sum_{j=1}^{m}[-\frac{\Phi_j\int\psi^-_2\Psi_j\mathrm{d}x}{(\psi^-_2)^2}(a_4+\int\psi^-_2\psi^+\mathrm{d}x)+\frac{\int\psi^+\Phi_j\mathrm{d}x\int\psi^-_2\Psi_j\mathrm{d}x}{\psi^-_2}]
\nonumber\\
&=&4\sum_{j=1}^{m}\frac{\int\Psi_j\psi^-_2\mathrm{d}x}{\psi^-_2}[-\frac{\Phi_j}{\psi^-_2}(a_4+\int\psi^-_2\psi^+\mathrm{d}x)+\int\psi^+\Phi_j\mathrm{d}x]
\nonumber\\
&=&-4\sum_{j=1}^{m}\frac{\int\Psi_j\psi^-_2\mathrm{d}x}{\psi^-_2}\{\int[(a_4+\int\psi^-_2\psi^+\mathrm{d}x)d(\frac{\Phi_j}{\psi^-_2})]\mathrm{d}x\}
\nonumber\\
&=&-4\sum_{j=1}^{m}\frac{\int\Psi_j\psi^-_2\mathrm{d}x}{\psi^-_2}\int[\frac{(a_4+\int\psi^-_2\psi^+\mathrm{d}x)}{\psi^-_2}\frac{(\Phi_{j,x}\psi^-_2-\Phi_j\psi^-_{2,x})}{\psi^-_2}]\mathrm{d}x
\nonumber\\
&=&4\sum_{j=1}^{m}\Psi_j[-1]\int\psi^+[-1]\Phi_j[-1]\mathrm{d}x
\nonumber\\
&=&\text{the r.h.s. of (\ref{c6})}.
\end{eqnarray}
This completes the proof.\\

From Theorem 3.1 and Theorem 3.2, we can obtain the binary Darboux
transformation for the system (\ref{b2}) by choosing $a_4=0$,\
$\psi^+_1[-1]=\frac{C+\int\psi^+_1\psi^-_2\mathrm{d}x}{\psi^-_2}$\
($C$ is a constant) as follows
\begin{subequations}
\label{c8}
\begin{equation}
\label{c81}
\psi^+[-1,+1]=\psi^+-\frac{\psi^+_1\int\psi^-_2\psi^+\mathrm{d}x}{C+\int\psi^+_1\psi^-_2\mathrm{d}x},
\end{equation}
\begin{equation}
\label{c82}
u[-1,+1]=u+2\partial^2\mathrm{ln}(C+\int\psi^-_2\psi^+_1\mathrm{d}x),
\end{equation}
\begin{equation}
\label{c83}
 \Phi_j[-1,+1]=\Phi_j-\frac{\psi^-_2\int\psi^+_1\Phi_j\mathrm{d}x}{C+\int\psi^-_2\psi^+_1\mathrm{d}x},
\end{equation}
\begin{equation}
\label{c84}
 \Psi_j[-1,+1]=\Psi_j-\frac{\psi^+_1\int\psi^-_2\Psi_j\mathrm{d}x}{C+\int\psi^-_2\psi^+_1\mathrm{d}x},\qquad
 j=1,...,m.
\end{equation}
\end{subequations}

Substituting (\ref{c8}) into equation (\ref{b22}) gives the
equality
\begin{equation}
\label{c9}
\psi^+[-1,+1]_t=A^+(u[-1,+1])\psi^+[-1,+1]+T^+_m(\Psi[-1,+1],\Phi[-1,+1])\psi^+[-1,+1].
\end{equation}

Both sides of the equality (\ref{c9}) are polynomials w.r.t. the
term $(C+\int\psi^-_2\psi^+_1\mathrm{d}x)^{-1}$. For example, the
left hand side of (\ref{c9}) is a polynomial of order 2 w.r.t.
$(C+\int\psi^-_2\psi^+_1\mathrm{d}x)^{-1}$ as follows
\begin{equation}
\label{c10}
\begin{array}{rcl}
\psi^+[-1,+1]_t&=&[\psi^+-\frac{\psi^+_1\int\psi^-_2\psi^+\mathrm{d}x}{C+\int\psi^+_1\psi^-_2\mathrm{d}x}]_t
\\
&=&\psi^+_t-\frac{\psi^+_{1,t}(\int\psi^-_2\psi^+\mathrm{d}x)}{C+\int\psi^+_1\psi^-_2\mathrm{d}x}-\frac{\psi^+_1(\int\psi^-_2\psi^+\mathrm{d}x)_t}{C+\int\psi^+_1\psi^-_2\mathrm{d}x}+\frac{\psi^+_1(\int\psi^-_2\psi^+\mathrm{d}x)(\int\psi^-_2\psi^+\mathrm{d}x)_t}{(C+\int\psi^+_1\psi^-_2\mathrm{d}x)^2}.
\\
&\equiv&\sum_{j=0}^2L_j[C+\int\psi^-_2\psi^+_1\mathrm{d}x]^{-j}.
\end{array}
\end{equation}

By a tedious computation, the right hand side of (\ref{c9}) is
displayed to be a polynomial of the order less than 4 w.r.t.
$(C+\int\psi^-_2\psi^+_1\mathrm{d}x)^{-1}$. We denote it by
$\sum_{j=0}^4R_j[C+\int\psi^-_2\psi^+_1\mathrm{d}x]^{-j}$.
 Since (\ref{c9}) holds for arbitrary constant $C$ and $L_j,R_j$ never contain $C$,
 we have $$L_j=R_j,\  j=0,1,2\ \   and \ \ R_j=0,\  j=3,4.$$

 If we replace $C$ in (\ref{c8}) by $C(t)$, an arbitrary function in $t$, and substitute equations
(\ref{c8}) into both sides of the equations of (\ref{b2}) again,
we will find that (\ref{b21}) is also covariant w.r.t. (\ref{c8}).
The right hand side of (\ref{c9}) turns to be
$\sum_{j=0}^2R_j[C(t)+\int\psi^-_2\psi^+_1\mathrm{d}x]^{-j}$,  but
the left hand side of (\ref{c9}) does not equal to
$\sum_{j=0}^2L_j[C(t)+\int\psi^-_2\psi^+_1\mathrm{d}x]^{-j}$ any
more. In other words, (\ref{b22}) does not covariant w.r.t.
(\ref{c8}) any longer when $C$ replaced by $C(t)$. In fact we have
the following theorem.

\begin{theorem}
 Given $u$,$\Psi_1,...,\Psi_m,\Phi_1,...,\Phi_m$ a solution of
the KPESCS (\ref{b1}) and let $\psi^+_1$ and $\psi^-_2$ be
solutions of the system (\ref{b2}) and (\ref{b3}) respectively,
then the transformation with $C(t)$, an arbitrary function in $t$
defined by
\begin{subequations}
\label{d1}
\begin{equation}
\label{d11}
\psi^+[-1,+1]=\psi^+-\frac{\psi^+_1\int\psi^-_2\psi^+\mathrm{d}x}{C(t)+\int\psi^+_1\psi^-_2\mathrm{d}x},
\end{equation}
\begin{equation}
\label{d12}
u[-1,+1]=u+2\partial^2\mathrm{ln}(C(t)+\int\psi^-_2\psi^+_1\mathrm{d}x),
\end{equation}
\begin{equation}
\label{d13}
 \Phi_j[-1,+1]=\Phi_j-\frac{\psi^-_2\int\psi^+_1\Phi_j\mathrm{d}x}{C(t)+\int\psi^-_2\psi^+_1\mathrm{d}x},
\end{equation}
\begin{equation}
\label{d14}
 \Psi_j[-1,+1]=\Psi_j-\frac{\psi^+_1\int\psi^-_2\Psi_j\mathrm{d}x}{C(t)+\int\psi^-_2\psi^+_1\mathrm{d}x},\qquad j=1,...,m,
\end{equation}
\text{and}
\begin{equation}
\label{d15}
\Psi_{m+1}[-1,+1]=\frac{1}{2}\frac{\sqrt{\dot{C}(t)}\psi^+_1}{C(t)+\int\psi^-_2\psi^+_1\mathrm{d}x},\
\
\Phi_{m+1}[-1,+1]=\frac{1}{2}\frac{\sqrt{\dot{C}(t)}\psi^-_2}{C(t)+\int\psi^-_2\psi^+_1\mathrm{d}x},
\end{equation}
\end{subequations}
transforms (\ref{b12}), (\ref{b13}) and (\ref{b2}) respectively
into
\begin{subequations}
\label{d2}
\begin{equation}
\label{d21}
    \alpha\Phi_j[-1,+1]_y=\Phi_j[-1,+1]_{xx}+u[-1,+1]\Phi_j[-1,+1],
\end{equation}
\begin{equation}
\label{d22}
    \alpha\Psi_j[-1,+1]_y=-\Psi_j[-1,+1]_{xx}-u[-1,+1]\Psi_j[-1,+1],\ \ j=1,...,m+1,
\end{equation}
\begin{equation}
\label{d23}
    \alpha\psi^+[-1,+1]_y=-\psi^+[-1,+1]_{xx}-u[-1,+1]\psi^+[-1,+1],
\end{equation}
\begin{equation}
\label{d24}
    \psi^+[-1,+1]_t=A^+(u[-1,+1])\psi^+[-1,+1]+T^+_{m+1}(\Psi[-1,+1],\Phi[-1,+1])\psi^+[-1,+1].
\end{equation}
\end{subequations}
So $u[-1,+1]$, $\Psi_1[-1,+1]$, ..., $\Psi_{m+1}[-1,+1]$,
$\Phi_1[-1,+1]$, ..., $\Phi_{m+1}[-1,+1]$ is a solution of the
KPESCS (\ref{b1}) with degree $m+1$.
\end{theorem}
{\bf{Proof}}:\ Equations (\ref{d21}), (\ref{d22}) and (\ref{d23})
hold obviously. We only need to prove (\ref{d24}). Substituting
(\ref{d11}) into the left hand side of (\ref{d24}), we have
\begin{equation}
\label{d4}
\begin{array}{rcl}
\psi^+[-1,+1]_t&=&[\psi^+-\frac{\psi^+_1\int\psi^-_2\psi^+\mathrm{d}x}{C(t)+\int\psi^+_1\psi^-_2\mathrm{d}x}]_t
\\
&=&\sum_{j=0}^2L_j[C(t)+\int\psi^-_2\psi^+_1\mathrm{d}x]^{-j}+\frac{\dot{C}(t)\psi^+_1\int\psi^+\psi^-_2\mathrm{d}x}{(C(t)+\int\psi^+_1\psi^-_2\mathrm{d}x)^2}
\\
&=&\sum_{j=0}^2R_j[C(t)+\int\psi^-_2\psi^+_1\mathrm{d}x]^{-j}+\frac{\dot{C}(t)\psi^+_1\int\psi^+\psi^-_2\mathrm{d}x}{(C(t)+\int\psi^+_1\psi^-_2\mathrm{d}x)^2}
\\
&=&A^+(u[-1,+1])\psi^+[-1,+1]+T^+_m(\Psi[-1,+1],\Phi[-1,+1])\psi^+[-1,+1]
\\&&+\frac{\dot{C}(t)\psi^+_1\int\psi^+\psi^-_2\mathrm{d}x}{(C(t)+\int\psi^+_1\psi^-_2\mathrm{d}x)^2}.
\end{array}
\end{equation}

So we only need to prove
\begin{equation}
\label{d5}
 4\Psi_{m+1}[-1,+1]\int\Phi_{m+1}[-1,+1]\psi^+[-1,+1]\mathrm{d}x=\frac{\dot{C}(t)\psi^+_1\int\psi^+\psi^-_2\mathrm{d}x}{(C(t)+\int\psi^+_1\psi^-_2\mathrm{d}x)^2}.
\end{equation}

In fact, we have
\begin{equation}
\label{d6}
\begin{array}{rcl}
\text{the l.h.s of (\ref{d5})} &=&
4\times\frac{1}{2}\frac{\sqrt{\dot{C}(t)}\psi^+_1}{C(t)+\int\psi^+_1\psi^-_2\mathrm{d}x}\int[(\psi^+-\frac{\psi^+_1\int\psi^+\psi^-_2\mathrm{d}x}{C(t)+\int\psi^+_1\psi^-_2\mathrm{d}x})\times\frac{1}{2}\frac{\sqrt{\dot{C}(t)}\psi^-_2}{C(t)+\int\psi^-_2\psi^+_1\mathrm{d}x}]\mathrm{d}x
\\
&=&\frac{\dot{C}(t)\psi^+_1}{C(t)+\int\psi^+_1\psi^-_2\mathrm{d}x}\int[(\psi^+-\frac{\psi^+_1\int\psi^+\psi^-_2\mathrm{d}x}{C(t)+\int\psi^+_1\psi^-_2\mathrm{d}x})\frac{\psi^-_2}{C(t)+\int\psi^+_1\psi^-_2\mathrm{d}x}]\mathrm{d}x
\\
&=&\frac{\dot{C}(t)\psi^+_1}{C(t)+\int\psi^+_1\psi^-_2\mathrm{d}x}\int\frac{\psi^+\psi^-_2(C(t)+\int\psi^+_1\psi^-_2\mathrm{d}x)-\psi^+_1\psi^-_2(\int\psi^+\psi^-_2\mathrm{d}x)}{(C(t)+\int\psi^+_1\psi^-_2\mathrm{d}x)^2}\mathrm{d}x,
\\
&=&\frac{\dot{C}(t)\psi^+_1}{C(t)+\int\psi^+_1\psi^-_2\mathrm{d}x}\int
d(\frac{\int\psi^+\psi^-_2\mathrm{d}x}{C(t)+\int\psi^+_1\psi^-_2\mathrm{d}x})
\\
&=&\frac{\dot{C}(t)\psi^+_1}{C(t)+\int\psi^+_1\psi^-_2\mathrm{d}x}\frac{\int\psi^+\psi^-_2\mathrm{d}x}{C(t)+\int\psi^+_1\psi^-_2\mathrm{d}x}
\\
&=&\text{the r.h.s of (\ref{d5})}
\end{array}
\end{equation}
This completes the proof.
\\
{\bf{Remark}}:

(1). For system (\ref{b3}), the DTs described in this section can
also be constructed, we omit the results here.

(2). (\ref{d12})-(\ref{d15}) offer us a non-auto-B\"{a}cklund
transformation between two KPESCSs with degrees $m$ and $m+1$
respectively. This DT can be used to construct solutions of
the KPESCS (\ref{b1}).\\
{\bf{Example 1}}: single-soliton solution of KPI equation with
self-consistent sources (KPIESCS).

If we set $\alpha=i$ in equation (\ref{b1}), we get the KPIESCS
\cite{Mel'nikov89(2)}
\begin{subequations}
\label{d7}
\begin{equation}
\label{d71}
     [u_t+6uu_x+u_{xxx}+8\sum_{j=1}^{m}(\Phi_j\Psi_j)_x]_x-3u_{yy} =0,
\end{equation}
\begin{equation}
\label{d72}
    i\Phi_{j,y}=\Phi_{j,xx}+u\Phi_j,
\end{equation}
\begin{equation}
\label{d73}
    i\Psi_{j,y}=-\Psi_{j,xx}-u\Psi_j,\qquad j=1,...,m.
\end{equation}
\end{subequations}
We take $u=0$ as the initial solution of (\ref{d7}) with $m=0$ and
let
\begin{subequations}
\label{d8}
\begin{equation}
\label{d81}
     \psi^+_1=e^{k_1x+ik^2_1y-4k^3_1t},\ \psi^-_2=e^{k_2x-ik^2_2y-4k^3_2t},\
     k_1=\mu+i\nu,\
     k_2=\mu-i\nu,\  \mu,\nu\in\mathbb{R},
\end{equation}
\begin{equation}
\label{d82}
    C(t)=\frac{1}{k_1+k_2}e^{2\beta(t)}=\frac{1}{2\mu} e^{2\beta(t)},
\end{equation}
\end{subequations}
where $\beta(t)$ is an arbitrary function in $t$. Then
$\psi^+_1=e^{\omega+\theta}$,\ $\psi^-_2=e^{-\omega+\theta}$ where
$$\omega=-i\nu x+i(\mu^2-\nu^2)y+4i(3\mu^2\nu-\nu^3)t,\ \ \theta=\mu
x+2\mu\nu y-4(\mu^3-3\mu\nu^2)t.$$

The single-soliton solution of the KPIESCS (\ref{d7}) with $m=1$
is given by
\begin{subequations}
\label{d9}
\begin{equation}
\label{d91}
\begin{array}{l}
     u[-1,+1]=2\partial^2\mathrm{ln}(C(t)+\int\psi^+_1\psi^-_2\mathrm{d}x)=2\partial^2\mathrm{ln}(\frac{1}{2\mu} e^{2\beta(t)}+\frac{1}{2\mu}e^{2\theta(t)})=2\mu^2\mathrm{sech}^2(\theta-\beta(t)),
\end{array}
\end{equation}
\begin{equation}
\label{d92}
\begin{array}{l}
    \Phi_1[-1,+1]=\frac{1}{2}\frac{\sqrt{\dot{C}(t)}\psi^-_2}{C(t)+\frac{1}{2\mu}e^{2\theta(t)}}=ae^{-\omega}\mathrm{sech}(\theta-\beta(t)),\ a=\frac{1}{2}\sqrt{\mu\dot{\beta}(t)}
\end{array}
\end{equation}
\begin{equation}
\label{d93}
\begin{array}{l}
    \Psi_1[-1,+1]=\overline{\Phi_1[-1,+1]}=ae^{\omega}\mathrm{sech}(\theta-\beta(t)).
\end{array}
\end{equation}
\end{subequations}

We can compare the results above with those in
\cite{Mel'nikov89(2)}.

{\bf{Case 1}}: If we set $$\theta-\beta(t)=\mu x+2\mu\nu
y-4(\mu^3-3\mu\nu^2)t-\beta(t)=\mu(x+2\nu y-\tau t),$$ then
$$\beta(t)=[\tau-4(\mu^2-3\nu^2)]\mu t,$$$$\dot{\beta}(t)=\mu[\tau-4(\mu^2-3\nu^2)],$$
then we have the following identity
\begin{equation}
\label{d10}
    a^2=\frac{1}{4}\mu\dot{\beta}(t)=\frac{\mu^2}{4}[\tau-4(\mu^2-3\nu^2)],
\end{equation}
which gives the relation of the parameters $a,\ \tau,\ \mu,\ \nu$
appearing in the solution (\ref{d9}). The solution (\ref{d9})
together with (\ref{d10}) is also obtained in
\cite{Mel'nikov89(2)}.

{\bf{Case 2}}: More generally, if we set
$$\theta-\beta(t)=\mu(x+2\nu y-f(t)),$$ then
$$f(t)=4(\mu^2-3\nu^2)t+\frac{\beta(t)}{\mu},$$
then the relation identity of the parameters is
\begin{equation}
\label{e1}
    \frac{df}{dt}(t)=4(\mu^2-3\nu^2)+\frac{\dot{\beta}(t)}{\mu}=4(\mu^2-3\nu^2)+4\mu^{-2}a^2,
\end{equation}
which is also given in \cite{Mel'nikov89(2)}.

From {\bf{Case 1}} and {\bf{Case 2}}, we see that the solutions
obtained through the generalized binary DT are consistent with
those in \cite{Mel'nikov89(2)}, and the method used here is more
natural and simple.\\
{\bf{Example 2}}: lump solution (rational solution) of the
KPIESCS.

We take $u=0$,\ $\Phi_1=\Psi_1=1$ as the initial solution of
(\ref{d7}) with $m=1$ and let $$
\psi^+_1=(x+2iky-12k^2t-\frac{4}{k^2}t)e^{kx+ik^2y-4k^3t+\frac{4}{k}t},\
\psi^-_2=\overline{\psi^+_1},\  k\in\mathbb{C}.$$ Set
$k=\mu+i\nu$,\ $\mu,\ \nu \in \mathbb{R}$,
then$$\psi^+_1=e^{\theta_1+i\theta_2}(\alpha_1+\alpha_2i),\ \
\psi^-_2=e^{\theta_1-i\theta_2}(\alpha_1-\alpha_2i),$$ where
$$\theta_1=\mu x-2\mu\nu y-
4\mu(\mu^2-3\nu^2)t+\frac{4\mu}{\mu^2+\nu^2}t,\ \theta_2=\nu
x+(\mu^2-\nu^2)y-4\nu(3\mu^2-\nu^2)t-\frac{4\nu}{\mu^2+\nu^2}t,$$$$\alpha_1=x-2\nu
y-12(\mu^2-\nu^2)t-4\frac{\mu^2-\nu^2}{(\mu^2+\nu^2)^2}t ,\
\alpha_2=2\mu y-24\mu\nu t+\frac{8\mu\nu}{(\mu^2+\nu^2)^2}t.$$ By
DT (\ref{d1}) with $C(t)\equiv 0$, we get the lump solution of the
KPIESCS (\ref{d7}) with $m=1$ as follows
\begin{subequations}
\label{e2}
\begin{equation}
\label{e21}
u[-1,+1]=2\partial^2\mathrm{ln}\int\psi^+_1\psi^-_2\mathrm{d}x=2\partial^2\mathrm{ln}[\alpha^2_2+(\alpha_1-\frac{1}{2\mu})^2+\frac{1}{4\mu^2}],
\end{equation}
\begin{equation}
\label{e22}
\begin{array}{lll}
\Phi_1[-1,+1]&=&\Phi_1-\frac{\psi^-_2\int\psi^+_1\Phi_1\mathrm{d}x}{\int\psi^+_1\psi^-_2\mathrm{d}x}
    \\&=&1-2\mu\frac{(\alpha^2_1+\alpha^2_2)\mu(\mu^2+\nu^2)-\alpha_1(\mu^2-\nu^2)+2\mu\nu\alpha_2+[-(\alpha^2_1+\alpha^2_2)(\nu\mu^2+\nu^3)+(\mu^2-\nu^2)\alpha_2+2\mu\nu\alpha_1]i}{(\mu^2+\nu^2)^2[\alpha^2_2+(\alpha_1-\frac{1}{2\mu})^2+\frac{1}{4\mu^2}]},
\end{array}
\end{equation}
\begin{equation}
\label{e23}
\begin{array}{lll}
\Psi_1[-1,+1]&=&\overline{\Phi_1[-1,+1]}\\
    &=&1-2\mu\frac{(\alpha^2_1+\alpha^2_2)\mu(\mu^2+\nu^2)-\alpha_1(\mu^2-\nu^2)+2\mu\nu\alpha_2-[-(\alpha^2_1+\alpha^2_2)(\nu\mu^2+\nu^3)+(\mu^2-\nu^2)\alpha_2+2\mu\nu\alpha_1]i}{(\mu^2+\nu^2)^2[\alpha^2_2+(\alpha_1-\frac{1}{2\mu})^2+\frac{1}{4\mu^2}]}.
\end{array}
\end{equation}
\end{subequations}
{\bf{Example 3}}: single-soliton solution of the KPII equation
with self-consistent sources (KPIIESCS).

If we set $\alpha=1$ in equation (\ref{b1}), we get the KPIIESCS
\cite{Zhangdajun2003}
\begin{subequations}
\label{e3}
\begin{equation}
\label{e31}
     [u_t+6uu_x+u_{xxx}+8\sum_{j=1}^{m}(\Phi_j\Psi_j)_x]_x+3u_{yy} =0,
\end{equation}
\begin{equation}
\label{e32}
    \Phi_{j,y}=\Phi_{j,xx}+u\Phi_j,
\end{equation}
\begin{equation}
\label{e33}
    \Psi_{j,y}=-\Psi_{j,xx}-u\Psi_j,\qquad j=1,...,m.
\end{equation}
\end{subequations}
We take $u=0$ as the initial solution of (\ref{e3}) with $m=0$ and
let
\begin{subequations}
\label{e4}
\begin{equation}
\label{e41}
\begin{array}{l}
     \psi^+_1=e^{k_1x-k^2_1y-4k^3_1t}=e^{\xi_1},\ \
     \psi^-_2=e^{k_2x+k^2_2y-4k^3_2t}=e^{\xi_2},\ \ k_1, k_2 \in \mathbb{R},
\end{array}
\end{equation}
\begin{equation}
\label{e42}
    C(t)=\frac{1}{k_1+k_2}e^{2\beta(t)},
\end{equation}
\end{subequations}
where $\beta(t)$ is an arbitrary function in
    $t$. The single-soliton solution of the KPIIESCS (\ref{e3}) with $m=1$ is given by
\begin{eqnarray}
\label{e5}
u[-1,+1]&=&2\partial^2\mathrm{ln}(C(t)+\int\psi^+_1\psi^-_2\mathrm{d}x)=2\partial^2\mathrm{ln}(1+e^{\xi_1+\xi_2-2\beta(t)}),
\nonumber\\
\Phi_1[-1,+1]&=&\frac{1}{2}\frac{\sqrt{\dot{C}(t)}\psi^-_2}{C(t)+\frac{1}{k_1+k_2}e^{\xi_1+\xi_2}}=\frac{\sqrt{2(k_1+k_2)\dot{\beta}}e^{\xi_2-\beta(t)}}{2(1+e^{\xi_1+\xi_2-2\beta(t)})},
\nonumber\\
\Psi_1[-1,+1]&=&\frac{1}{2}\frac{\sqrt{\dot{C}(t)}\psi^+_1}{C(t)+\frac{1}{k_1+k_2}e^{\xi_1+\xi_2}}=\frac{\sqrt{2(k_1+k_2)\dot{\beta}}e^{\xi_1-\beta(t)}}{2(1+e^{\xi_1+\xi_2-2\beta(t)})}.
\end{eqnarray}

\section{The N-times Repeated Generalized Binary Darboux transformation for the KPESCS}
\setcounter{equation}{0}\hskip\parindent Assuming $f_1,...,f_n$
are $n$ arbitrary solutions of (\ref{b2}), $g_1,...,g_n$ are $n$
arbitrary solutions of (\ref{b3}), $C_1(t),...,C_n(t)$ are $n$
arbitrary functions in $t$, we define the following Wronskians:
\begin{equation}
\label{e6}
\begin{array}{lll}
W_1(f_1,...,f_n;g_1,...,g_n;C_1(t),...,C_n(t))&=&det(U_{n\times
n}),\\W_2(f_1,...,f_n;g_1,...,g_{n-1};C_1(t),...,C_{n-1}(t))&=&det(V_{n\times
n}),\\W_3(f_1,...,f_{n-1};g_1,...,g_n;C_1(t),...,C_{n-1}(t))&=&det(X_{n\times
n}),
\end{array}
\end{equation}
where
\begin{subequations}
\label{e7}
\begin{equation}
\label{e71}
     U_{i,j}=\delta_{i,j}C_i(t)+\int g_if_j\mathrm{d}x,\ \ \  i,j=1,...,n,
\end{equation}
\begin{equation}
\label{e72} V_{i,j}=\delta_{i,j}C_i(t)+\int g_if_j\mathrm{d}x,\
i=1,...,n-1,\ j=1,...,n;\ \  V_{n,j}=f_j,\  j=1,...,n,
\end{equation}
\begin{equation}
\label{e73} X_{i,j}=\delta_{i,j}C_i(t)+\int g_jf_i\mathrm{d}x,\
i=1,...,n-1,\ j=1,...,n;\ \ X_{n,j}=g_j,\ j=1,...,n.
\end{equation}
\end{subequations}

\begin{lemma} Assume $\psi^+,\psi^+_1,..., \psi^+_N$ are solutions of (\ref{b2}) and $\psi^-,\psi^-_{N+1},..., \psi^-_{N+N}$ are solutions
of (\ref{b3}). For notation simplicity, we denote
$\psi^+_{l+k}[l]=\psi^+_{l+k}[-l,+l]$,\
$\psi^-_{N+l+k}[l]=\psi^-_{N+l+k}[-l,+l]$, $0\leq l\leq N$, $1\leq
k\leq N-l$. Then
\begin{subequations}
\label{e8}
\begin{equation}
\label{e81}
\begin{array}{ll}
&W_1(\psi^+_m[m-1],...,\psi^+_{m+k}[m-1];\psi^-_{N+m}[m-1],...,\psi^-_{N+m+k}[m-1];C_m(t),...,C_{m+k}(t))\\
    =&\frac{W_1(\psi^+_{m-1}[m-2],...,\psi^+_{m+k}[m-2];\psi^-_{N+m-1}[m-2],...,\psi^-_{N+m+k}[m-2];C_{m-1}(t),...,C_{m+k}(t))}{C_{m-1}(t)+\int\psi^+_{m-1}[m-2]\psi^-_{N+m-1}[m-2]\mathrm{d}x},
\end{array}
\end{equation}
\begin{equation}
\label{e82}
\begin{array}{ll}
&W_2(\psi^+_m[m-1],...,\psi^+_{m+k}[m-1],\psi^+[m-1];\psi^-_{N+m}[m-1],...,\psi^-_{N+m+k}[m-1];\\
&C_m(t),...,C_{m+k}(t))\\
=&\frac{W_2(\psi^+_{m-1}[m-2],...,\psi^+_{m+k}[m-2],\psi^+[m-2];\psi^-_{N+m-1}[m-2],...,\psi^-_{N+m+k}[m-2];C_{m-1}(t),...,C_{m+k}(t))}{C_{m-1}(t)+\int\psi^+_{m-1}[m-2]\psi^-_{N+m-1}[m-2]\mathrm{d}x},
\end{array}
\end{equation}
\begin{equation}
\label{e83}
\begin{array}{ll}
&W_3(\psi^+_m[m-1],...,\psi^+_{m+k}[m-1];\psi^-_{N+m}[m-1],...,\psi^-_{N+m+k}[m-1],\psi^-[m-1];\\
&C_m(t),...,C_{m+k}(t))\\
=&\frac{W_3(\psi^+_{m-1}[m-2],...,\psi^+_{m+k}[m-2];\psi^-_{N+m-1}[m-2],...,\psi^-_{N+m+k}[m-2],\psi^-[m-2];C_{m-1}(t),...,C_{m+k}(t))}{C_{m-1}(t)+\int\psi^+_{m-1}[m-2]\psi^-_{N+m-1}[m-2]\mathrm{d}x}.
\end{array}
\end{equation}
\end{subequations}
\end{lemma}
{\bf{Proof}}: According to (\ref{d1}), we have
\begin{subequations}
\label{e9}
\begin{equation}
\label{e91}
\begin{array}{l}
 \psi^+_{m-1+j}[m-1]= \psi^+_{m-1+j}[m-2]-\frac{\psi^+_{m-1}[m-2]\int\psi^-_{N+m-1}[m-2]\psi^+_{m-1+j}[m-2]\mathrm{d}x}{C_{m-1}(t)+\int\psi^+_{m-1}[m-2]\psi^-_{N+m-1}[m-2]\mathrm{d}x},
\end{array}
\end{equation}

\begin{equation}
\label{e92}
\begin{array}{l}
\psi^-_{N+m-1+i}[m-1]=
\psi^-_{N+m-1+i}[m-2]-\frac{\psi^-_{N+m-1}[m-2]\int\psi^+_{m-1}[m-2]\psi^-_{N+m-1+i}[m-2]\mathrm{d}x}{C_{m-1}(t)+\int\psi^+_{m-1}[m-2]\psi^-_{N+m-1}[m-2]\mathrm{d}x}.
\end{array}
\end{equation}
\end{subequations}

By definition (\ref{e71}) and equations (\ref{e9}), we have
\begin{equation}
\label{e10}
\begin{array}{ll}
 &W_1(\psi^+_m[m-1],...,\psi^+_{m+k}[m-1];\psi^-_{N+m}[m-1],...,\psi^-_{N+m+k}[m-1];C_m(t),...,C_{m+k}(t))\\
 =&det(U_{i,j}),\ \ 1\leq i,j\leq k+1,
 \end{array}
\end{equation}
\begin{equation}
\label{f1}
\begin{array}{ll}
        &U_{i,j}\\
       =&\delta_{i,j}C_{m-1+i}(t)+\int\psi^+_{m-1+j}[m-1]\psi^-_{N+m-1+i}[m-1]\mathrm{d}x
        \\
        =&\delta_{i,j}C_{m-1+i}(t)+\int\psi^+_{m-1+j}[m-2]\psi^-_{N+m-1+i}[m-2]\mathrm{d}x-\frac{1}{C_{m-1}(t)+\int\psi^+_{m-1}[m-2]\psi^-_{N+m-1}[m-2]\mathrm{d}x}
        \\
        &(\int\psi^+_{m-1+j}[m-2]\psi^-_{N+m-1}[m-2]\mathrm{d}x)(\int\psi^+_{m-1}[m-2]\psi^-_{N+m-1+i}[m-2]\mathrm{d}x)
        \\
        =&\delta_{i,j}C_{m-1+i}(t)+a_{i,j}-ba_{i,0}a_{0,j},
\end{array}
\end{equation}
where
$$a_{i,j}=\int\psi^+_{m-1+j}[m-2]\psi^-_{N+m-1+i}[m-2]\mathrm{d}x,\
\
b=\frac{1}{C_{m-1}(t)+\int\psi^+_{m-1}[m-2]\psi^-_{N+m-1}[m-2]\mathrm{d}x}.$$
Then
$$W_1(\psi^+_{m}[m-1],...,\psi^+_{m+k}[m-1];\psi^-_{N+m}[m-1],...,\psi^-_{N+m+k}[m-1];C_m(t),...,C_{m+k}(t))=$$
\[\begin{vmatrix}
C_m(t)+a_{1,1}-ba_{1,0}a_{0,1} &a_{1,2}-ba_{1,0}a_{0,2}& \cdots &a_{1,k+1}-ba_{1,0}a_{0,k+1} \\
a_{2,1}-ba_{2,0}a_{0,1} &C_{m+1}(t)+a_{2,2}-ba_{2,0}a_{0,2}& \cdots &a_{2,k+1}-ba_{2,0}a_{0,k+1} \\
\vdots&\vdots&\ddots&\vdots\\
a_{k+1,1}-ba_{k+1,0}a_{0,1} &a_{k+1,2}-ba_{k+1,0}a_{0,2}& \cdots
&C_{m+k}(t)+a_{k+1,k+1}-ba_{k+1,0}a_{0,k+1}
\end{vmatrix}\]
 \[=\begin{vmatrix}
    C_m(t)+a_{1,1} &a_{1,2}&\cdots &a_{1,k+1} \\
a_{2,1}&C_{m+1}(t)+a_{2,2}&\cdots&a_{2,k+1} \\
\vdots&\vdots&\ddots&\vdots\\
a_{k+1,1}&a_{k+1,2}&\cdots &C_{m+k}(t)+a_{k+1,k+1}
\end{vmatrix}\]
\[-ba_{0,1} \begin{vmatrix}
    a_{1,0} &a_{1,2}&\cdots &a_{1,k+1} \\
a_{2,0}&C_{m+1}(t)+a_{2,2}&\cdots&a_{2,k+1} \\
\vdots&\vdots&\ddots&\vdots\\
a_{k+1,0}&a_{k+1,2}& \cdots &C_{m+k}(t)+a_{k+1,k+1} \\
\end{vmatrix}\]
\[-ba_{0,2} \begin{vmatrix}
    C_m(t)+a_{1,1} &a_{1,0}&\cdots &a_{1,k+1} \\
a_{2,1}&a_{2,0}&\cdots&a_{2,k+1} \\
\vdots&\vdots&\ddots&\vdots\\
a_{k+1,1}&a_{k+1,0}&\cdots &C_{m+k}(t)+a_{k+1,k+1}
\end{vmatrix}\]
\[-\cdots-ba_{0,k+1} \begin{vmatrix}
    C_m(t)+a_{1,1} &a_{1,2}&\cdots&a_{1,k} &a_{1,0} \\
a_{2,1}&C_{m+1}(t)+a_{2,2}&\cdots &a_{2,k}&a_{2,0} \\
\vdots&\vdots&\ddots&\vdots&\vdots\\
a_{k+1,1}&a_{k+1,2}&\cdots &a_{k+1,k}&a_{k+1,0}
\end{vmatrix}\]
\[=b \begin{vmatrix}
    C_{m-1}(t)+a_{0,0} &a_{0,1}&a_{0,2}&\cdots &a_{0,k+1} \\
a_{1,0}&C_m(t)+a_{1,1}&a_{1,2}&\cdots &a_{1,k+1} \\
\vdots&\vdots&\vdots&\ddots&\vdots\\
a_{k+1,0}&a_{k+1,1}&a_{k+1,2}&\cdots &C_{m+k}(t)+a_{k+1,k+1}
\end{vmatrix}=\]
$$\frac
{W_1(\psi^+_{m-1}[m-2],...,\psi^+_{m+k}[m-2];\psi^-_{N+m-1}[m-2],...,\psi^-_{N+m+k}[m-2];C_{m-1}(t),...,C_{m+k}(t))}{C_{m-1}(t)+\int\psi^+_{m-1}[m-2]\psi^-_{N+m-1}[m-2]\mathrm{d}x}.$$
So the formula (\ref{e81}) holds. The formulas (\ref{e82}) and
(\ref{e83}) can be proved similarly. \\This completes the proof.
\begin{theorem}
Assume that $u, \Psi_1,\cdots,\Psi_m,\Phi_1,\cdots,\Phi_m$ is a
solution of the KPESCS (\ref{b1}), $\psi^+_1,...,\psi^+_N$ and
$\psi^-_{N+1},...,\psi^-_{N+N}$ are solutions of (\ref{b2}) and
(\ref{b3}) respectively, $C_1(t),...,C_N(t)$ are $N$ arbitrary
functions in $t$. Then the $N$-times repeated generalized binary
Darboux transformation for (\ref{b2}) is given by
\begin{subequations}
\label{f2}
\begin{equation}
\label{f21} \psi^+[-N,+N]=\frac {W_2(\psi^+_1,..., \psi^+_N,
\psi^+;\psi^-_{N+1},...,
\psi^-_{N+N};C_1(t),...,C_N(t))}{W_1(\psi^+_1,...,
\psi^+_N;\psi^-_{N+1},..., \psi^-_{N+N};C_1(t),...,C_N(t))},
\end{equation}
\begin{equation}
\label{f22} u[-N,+N]=u+2\partial^2\mathrm{ln} W_1(\psi^+_1,...,
\psi^+_N;\psi^-_{N+1},..., \psi^-_{N+N};C_1(t),...,C_N(t)),
\end{equation}
\begin{equation}
\label{f23}
    \Psi_l[-N,+N]=-\frac {W_2(\psi^+_1,..., \psi^+_N,
\Psi_l;\psi^-_{N+1},...,
\psi^-_{N+N};C_1(t),...,C_N(t))}{W_1(\psi^+_1,...,
\psi^+_N;\psi^-_{N+1},..., \psi^-_{N+N};C_1(t),...,C_N(t))},
\end{equation}
\begin{equation}
\label{f24}
    \Phi_l[-N,+N]=-\frac {W_3(\psi^+_1,..., \psi^+_N;
\psi^-_{N+1},...,
\psi^-_{N+N},\Phi_l;C_1(t),...,C_N(t))}{W_1(\psi^+_1,...,
\psi^+_N;\psi^-_{N+1},..., \psi^-_{N+N};C_1(t),...,C_N(t))}, \quad
l=1,\cdots,m,
\end{equation}
\begin{equation}
\label{f25}
\begin{array}{l}
\Psi_{m+j}[-N,+N]\\
    =\frac{1}{2}\sqrt{\dot{C}_j(t)}\frac {W_2(\psi^+_1,...,\psi^+_{j-1},\psi^+_{j+1},..., \psi^+_N,\psi^+_j;
\psi^-_{N+1},...,\psi^-_{N+j-1},\psi^-_{N+j+1},...,
\psi^-_{N+N};C_1(t),...,C_{j-1}(t),C_{j+1}(t),...,C_N(t))}{W_1(\psi^+_1,...,
\psi^+_N;\psi^-_{N+1},..., \psi^-_{N+N};C_1(t),...,C_N(t))},
 \end{array}
\end{equation}
\begin{equation}
\label{f26}
\begin{array}{l}
    \Phi_{m+j}[-N,+N] \\
    =\frac{1}{2}\sqrt{\dot{C}_j(t)}\frac {W_3(\psi^+_1,...,\psi^+_{j-1},\psi^+_{j+1},..., \psi^+_N;
\psi^-_{N+1},...,\psi^-_{N+j-1},\psi^-_{N+j+1},...,
\psi^-_{N+N},\psi^-_{N+j};C_1(t),...,C_{j-1}(t),C_{j+1}(t),...,C_N(t))}{W_1(\psi^+_1,...,
\psi^+_N;\psi^-_{N+1},..., \psi^-_{N+N};C_1(t),...,C_N(t))},
\\ \quad j=1,\cdots,N.
 \end{array}
\end{equation}
\end{subequations}
Namely,
\begin{subequations}
\label{f3}
\begin{equation}
\label{f31}
    \alpha\Psi^+_j[-N,+N]_y=-\Psi^+_j[-N,+N]_{xx}-u[-N,+N]\Psi^+_j[-N,+N],
\end{equation}
\begin{equation}
\label{f32}
    \alpha\Phi^-_j[-N,+N]_y=\Phi^-_j[-N,+N]_{xx}+u[-N,+N]\Phi^-_j[-N,+N],\
    \ j=1,...,m+N,
\end{equation}
\begin{equation}
\label{f33}
    \alpha\psi^+[-N,+N]_y=-\psi^+[-N,+N]_{xx}-u[-N,+N]\psi^+[-N,+N],
\end{equation}
\begin{equation}
\label{f34}
    \psi^+[-N,+N]_t=A^+(u[-N,+N])\psi^+[-N,+N]+T^+_{m+N}(\Psi[-N,+N],\Phi[-N,+N])\psi^+[-N,+N].
\end{equation}
\end{subequations}
So $u[-N,+N]$,\ $\Psi_j[-N,+N],\ \Phi_j[-N,+N],\ j=1,...,m+N$
satisfy the KPESCS (\ref{b1}) with degree $(m+N)$.
\end{theorem}
{\bf{Proof}}: For notation simplicity, we denote $\psi^+[-l,+l]$,
$u[-l,+l]$, $\Psi_i[-l,+l]$, $\Phi_i[-l,+l]$ by $\psi^+[l]$,
$u[l]$, $\Psi_i[l]$, $\Phi_i[l]$ respectively, $0\leq l\leq N,
1\leq i\leq m+N$. By (\ref{d1}) and (\ref{e8}),
\begin{equation}
\label{f4}
\begin{array}{lll}
    \psi^+[N]&=&\psi^+[N-1]-\frac{\psi^+_N[N-1]\int\psi^-_{N+N}[N-1]\psi^+[N-1]\mathrm{d}x}{C_N(t)+\int\psi^-_{N+N}[N-1]\psi^+_N[N-1]\mathrm{d}x}
    \nonumber\\
    &=&\frac{W_2(\psi^+_{N}[N-1],\psi^+[N-1];\psi^-_{N+N}[N-1];C_N(t))}{W_1(\psi^+_N[N-1];\psi^-_{N+N}[N-1];C_N(t))}
    \nonumber\\
    &=&\frac{W_2(\psi^+_{N-1}[N-2],\psi^+_N[N-2],\psi^+[N-2];\psi^-_{N+N-1}[N-2],\psi^-_{N+N}[N-2];C_{N-1}(t),C_N(t))}{C_{N-1}(t)+\int\psi^-_{N+N-1}[N-2]\psi^+_{N-1}[N-2]\mathrm{d}x}
    \nonumber\\
    &&\times \frac{C_{N-1}(t)+\int\psi^-_{N+N-1}[N-2]\psi^+_{N-1}[N-2]\mathrm{d}x}{W_1(\psi^+_{N-1}[N-2],\psi^+_N[N-2];\psi^-_{N+N-1}[N-2],\psi^-_{N+N}[N-2];C_{N-1}(t),C_N(t))}
    \nonumber\\
    &=&\cdots
    \nonumber\\
    &=&\frac {W_2(\psi^+_1,..., \psi^+_N,
\psi^+;\psi^-_{N+1},...,
\psi^-_{N+N};C_1(t),...,C_N(t))}{W_1(\psi^+_1,...,
\psi^+_N;\psi^-_{N+1},..., \psi^-_{N+N};C_1(t),...,C_N(t))},
 \end{array}
\end{equation}
so (\ref{f21}) holds, Similarly we can prove (\ref{f23}) and
(\ref{f24}) hold,
\begin{equation}
\label{f5}
\begin{array}{lll}
    u[N]&=&u[N-1]+2\partial^2\mathrm{ln}(C_N(t)+\int\psi^+_N[N-1]\psi^-_{N+N}[N-1]\mathrm{d}x)
    \nonumber\\
    &=&u[N-1]+2\partial^2\mathrm{ln} W_1(\psi^+_N[N-1];\psi^-_{N+N}[N-1];C_N(t))
    \nonumber\\
    &=&u[N-2]+2\partial^2\mathrm{ln}(C_{N-1}(t)+\int
    \psi^+_{N-1}[N-2]\psi^-_{N+N-1}[N-2]\mathrm{d}x)+
    \nonumber\\
    &&2\partial^2\mathrm{ln}[\frac{W_1(\psi^+_{N-1}[N-2],\psi^+_N[N-2];\psi^-_{N+N-1}[N-2],\psi^-_{N+N}[N-2];C_{N-1}(t),C_N(t))}{C_{N-1}(t)
    +\int\psi^+_{N-1}[N-2]\psi^-_{N+N-1}[N-2]\mathrm{d}x}]
    \nonumber\\
    &=&u[N-2]+2\partial^2\mathrm{ln} W_1(\psi^+_{N-1}[N-2],\psi^+_N[N-2];\psi^-_{N+N-1}[N-2],\psi^-_{N+N}[N-2];
    \nonumber\\
    &&C_{N-1}(t),C_N(t))
    \nonumber\\
    &=&\cdots\\
    &=&u+2\partial^2\mathrm{ln} W_1(\psi^+_1,\cdots,\psi^+_N;\psi^-_{N+1},\cdots,\psi^-_{N+N};C_1(t),\cdots,C_N(t)),
\end{array}
\end{equation}
so (\ref{f22}) holds.\\
From (\ref{d11}) and  (\ref{d15}), we have
$$\psi^+_j[j]=\psi^+_j[j-1]-\frac{\psi^+_j[j-1]\int\psi^+_j[j-1]\psi^-_{N+j}[j-1]\mathrm{d}x}{C_j(t)+\int\psi^+_j[j-1]\psi^-_{N+j}[j-1]\mathrm{d}x}\\
=\frac{C_j(t)\psi^+_j[j-1]}{C_j(t)+\int\psi^+_j[j-1]\psi^-_{N+j}[j-1]\mathrm{d}x},$$
$$\Psi_{m+j}[j]=\frac{1}{2}\frac{\sqrt{\dot{C}_j(t)}\psi^+_j[j-1]}{C_j(t)+\int\psi^+_j[j-1]\psi^-_{N+j}[j-1]\mathrm{d}x},$$
so
$$\Psi_{m+j}[j]=\frac{1}{2}\frac{\sqrt{\dot{C}_j(t)}}{C_j(t)}\psi^+_j[j].$$
Analogously as above, we have
\begin{equation}
\label{f6}
\begin{array}{ll}
    &\Psi_{m+j}[N]\\
    =&\frac{W_2(\psi^+_N[N-1],\Psi_{m+j}[N-1];\psi^-_{N+N}[N-1];C_N(t))}{W_1(\psi^+_N[N-1];\psi^-_{N+N}[N-1];C_N(t))}
    \nonumber\\
    =&\frac{W_2 (\psi^+_{j+1}[j],\cdots,\psi^+_N[j],\Psi_{m+j}[j];\psi^-_{N+j+1}[j],\cdots,\psi^-_{N+N}[j];C_{j+1}(t),\cdots,C_N(t))}{W_1(\psi^+_{j+1}[j],\cdots,\psi^+_N[j];\psi^-_{N+j+1}[j],\cdots,\psi^-_{N+N}[j];C_{j+1}(t),\cdots,C_N(t))}
    \nonumber\\
    =&\frac{\sqrt{\dot{C}_j(t)}}{2C_j(t)}\frac{W_2 (\psi^+_{j+1}[j],\cdots,\psi^+_N[j],\psi^+_j[j];\psi^-_{N+j+1}[j],\cdots,\psi^-_{N+N}[j];C_{j+1}(t),\cdots,C_N(t))}{W_1(\psi^+_{j+1}[j],\cdots,\psi^+_N[j];\psi^-_{N+j+1}[j],\cdots,\psi^-_{N+N}[j];C_{j+1}(t),\cdots,C_N(t))}
    \nonumber\\
    =&\frac{\sqrt{\dot{C}_j(t)}}{2C_j(t)}\frac{W_2(\psi^+_1,\cdots,\psi^+_N,\psi^+_j;\psi^-_{N+1},\cdots,\psi^-_{N+N};C_1(t),\cdots,C_N(t))}{W_1(\psi^+_1,\cdots,\psi^+_N;\psi^-_{N+1},\cdots,\psi^-_{N+N};C_1(t),\cdots,C_{N}(t))}
    \\
    =&\frac{1}{2}\sqrt{\dot{C}_j(t)}\frac {W_2(\psi^+_1,...,\psi^+_{j-1},\psi^+_{j+1},..., \psi^+_N,\psi^+_j;
\psi^-_{N+1},...,\psi^-_{N+j-1},\psi^-_{N+j+1},...,
\psi^-_{N+N};C_1(t),...,C_{j-1}(t),C_{j+1}(t),...,C_N(t))}{W_1(\psi^+_1,...,
\psi^+_N;\psi^-_{N+1},..., \psi^-_{N+N};C_1(t),...,C_N(t))},
\end{array}
\end{equation}
so (\ref{f25}) holds. In a similar way we can prove (\ref{f26})
holds. This completes the proof.
\\
{\bf{Remark}}:

(1). If $C_1(t),\cdots,C_N(t)$ are replaced by $N$ arbitrary
constants $C_1,\dots,C_N$, then (\ref{f2}) gives the $N$-times
repeated Darboux transformation for (\ref{c8}).

(2). The $N$-times repeated generalized binary Darboux
transformation (\ref{f2}) provides a non-auto B\"{a}cklund
transformation between two KPESCSs with degrees $m$ and $(m+N)$
respectively. We now use this to construct some interesting
solutions for the KPESCS.
\\
{\bf{Example 4}}: (Degenerate case)

When some $C_j(t)$ in (\ref{f2}) are chosen to be arbitrary
numerical constants $C_j$, we can get the some interesting
solutions for the KPESCS (\ref{b1}) in the degenerate case
("degenerate" means the number of soliton is larger than the
degree of the equation). In the following, we will take the
KPIIESCS (\ref{e3}) for example and choose $u=0$ as the initial
solution of it with $m=0$.

{\bf{Example 4.1}}: two-soliton solution of the KPIIESCS.

Take
$$\psi^+_i=e^{\eta_i},\ \ \ \psi^-_{2+i}=e^{\theta_i},\ \ \ i=1,2, $$
$$\eta_i=l_ix-l^2_iy-4l^3_it,\ \ \ \theta_i=k_ix+k^2_iy-4k^3_it,\ \ \ l_i,k_i\in\mathbb{R},$$
$$C_1(t)=\frac{1}{k_1+l_1}e^{2\beta(t)},\ \ \ C_2(t)\equiv\frac{c}{k_2+l_2},$$
where $\beta(t)$ is an arbitrary function in $t$ and $c$ is an
arbitrary constant. From (\ref{f22}), (\ref{f25}) and (\ref{f26}),
we get the 2-soliton solution of the KPIIESCS (\ref{e3}) with
$m=1$ as follows
\begin{equation}
\label{f7}
\begin{array}{ll}
    &u(x,y,t)[-2,+2]\nonumber\\
    =&2\partial^2\mathrm{ln}W_1(\psi^+_1,\psi^+_2;\psi^-_3,\psi^-_4;C_1(t),C_2(t))\\
    =&2\partial^2\mathrm{ln}\left| \begin{array}{cc}
    C_1(t)+\int\psi^-_3\psi^+_1\mathrm{d}x & \int\psi^-_3\psi^+_2\mathrm{d}x\\
\int\psi^-_4\psi^+_1\mathrm{d}x &  C_2(t)+\int\psi^-_4\psi^+_2\mathrm{d}x\\
\end{array} \right|\\
   =&2\partial^2\mathrm{\mathrm{ln}}\left| \begin{array}{cc}
    \frac{1}{k_1+l_1}e^{2\beta(t)}+ \frac{1}{k_1+l_1}e^{\theta_1+\eta_1}& \frac{1}{k_1+l_2}e^{\theta_1+\eta_2}\\
\frac{1}{k_2+l_1}e^{\theta_2+\eta_1} &  \frac{c}{k_2+l_2}e^{\theta_2+\eta_2}\\
\end{array} \right|\\
   =&2\partial^2\mathrm{\mathrm{ln}}[\frac{ce^{2\beta(t)}}{(k_1+l_1)(k_2+l_2)}(1+e^{\theta_1+\eta_1-2\beta(t)}+e^{\theta_2+\eta_2+\xi_0}+
   \frac{(k_1-k_2)(l_1-l_2)}{(k_2+l_1)(k_1+l_2)}e^{\theta_1+\theta_2+\eta_1+\eta_2-2\beta(t)+\xi_0})]\\
   =&2\partial^2\mathrm{\mathrm{ln}}[1+e^{\theta_1+\eta_1-2\beta(t)}+e^{\theta_2+\eta_2+\xi_0}+
   \frac{(k_1-k_2)(l_1-l_2)}{(k_2+l_1)(k_1+l_2)}e^{\theta_1+\theta_2+\eta_1+\eta_2-2\beta(t)+\xi_0}],
\end{array}
\end{equation}
where $\xi_0=-\mathrm{\mathrm{ln}}c$,
\begin{equation}
\label{f8}
\begin{array}{ll}
    &\Psi_1[-2,+2]\\
    =&\frac{1}{2}\sqrt{\dot{C}_1(t)}\frac{W_2(\psi^+_2,\psi^+_1;\psi^-_4;C_2(t))}{W_1(\psi^+_1,\psi^+_2;\psi^-_3,\psi^-_4;C_1(t),C_2(t))}\nonumber\\
    =&\frac{1}{2}\sqrt{2(k_1+l_1)\dot{\beta}(t)}e^{\eta_1-\beta(t)}\frac{1+\frac{l_1-l_2}{k_2+l_1}e^{\theta_2+\eta_2+\xi_0}}{1+e^{\theta_1+\eta_1-2\beta(t)}+e^{\theta_2+\eta_2+\xi_0}+
   \frac{(k_1-k_2)(l_1-l_2)}{(k_2+l_1)(k_1+l_2)}e^{\theta_1+\theta_2+\eta_1+\eta_2-2\beta(t)+\xi_0}},
\end{array}
\end{equation}
\begin{equation}
\label{f9}
\begin{array}{ll}
    &\Phi_1[-2,+2]\\
    =&\frac{1}{2}\sqrt{\dot{C}_1(t)}\frac{W_3(\psi^+_2;\psi^-_4,\psi^-_3;C_2(t))}{W_1(\psi^+_1,\psi^+_2;\psi^-_3,\psi^-_4;C_1(t),C_2(t))}\nonumber\\
    =&\frac{1}{2}\sqrt{2(k_1+l_1)\dot{\beta}(t)}e^{\theta_1-\beta(t)}\frac{1+\frac{k_1-k_2}{k_1+l_2}e^{\theta_2+\eta_2+\xi_0}}{1+e^{\theta_1+\eta_1-2\beta(t)}+e^{\theta_2+\eta_2+\xi_0}+
   \frac{(k_1-k_2)(l_1-l_2)}{(k_2+l_1)(k_1+l_2)}e^{\theta_1+\theta_2+\eta_1+\eta_2-2\beta(t)+\xi_0}},
\end{array}
\end{equation}
$\Psi_2[-2,+2]=\Phi_2[-1,+2]=0$.

In a similar way, for $\forall N,m \in\mathbb{N}$,\  $N>m$, when
$C_j(t),j=1,...,m$ are taken to be arbitrary functions in $t$ and
$C_j(t),j=m+1,...,N$ are taken to be numerical constants, we can
get the $N$-soliton solution of (\ref{e3}) with degree $m$.

{\bf{Example 4.2}}: mixture of the exponential and rational
solutions of the KPIIESCS.

When take  $$\psi^+_1=e^{\zeta}(x-2qy-12q^2t),\ \
\psi^+_2=e^{\zeta},\ \ \zeta=qx-q^2y-4q^3t,\ \ q\in \mathbb{R}, $$
$$\psi^-_3=e^{\xi}(x+2ky-12k^2t),\ \
\psi^-_4=e^{\xi},\ \ \xi=kx+k^2y-4k^3t,\ \ k\in \mathbb{R}, $$
$$C_1(t)=\frac{1}{q+k}e^{2\beta(t)},\ \ \ C_2(t)\equiv\frac{c}{q+k},$$
where $\beta(t)$ is an arbitrary function in $t$ and $c$ is a
constant, we will get another solution (mixture of the exponential
and rational solutions) of the KPIIESCS with degree $m=1$ as
follows
\begin{equation}
    u(x,y,t)[-2,+2]=2\partial^2\mathrm{\mathrm{ln}}[1+e^{\zeta+\xi+\xi_0}+e^{\zeta+\xi-2\beta(t)}r+\frac{1}{(k+q)^2}e^{2(\zeta+\xi)-2\beta(t)+\xi_0}],\nonumber
\end{equation}
\begin{equation}
    \Psi_1[-2,+2]=\frac{1}{2}\sqrt{2(k+q)\dot{\beta}(t)}e^{\zeta-\beta(t)}\frac{x-2qy-12q^2t+\frac{1}{k+q}e^{\zeta+\xi+\xi_0}}{1+e^{\zeta+\xi+\xi_0}+e^{\zeta+\xi-2\beta(t)}r+\frac{1}{(k+q)^2}e^{2(\zeta+\xi)-2\beta(t)+\xi_0}},\nonumber
\end{equation}
\begin{equation}
\Phi_1[-2,+2]=\frac{1}{2}\sqrt{2(k+q)\dot{\beta}(t)}e^{\xi-\beta(t)}\frac{x+2ky-12k^2t+\frac{1}{k+q}e^{\zeta+\xi+\xi_0}}{1+e^{\zeta+\xi+\xi_0}+e^{\zeta+\xi-2\beta(t)}r+\frac{1}{(k+q)^2}e^{2(\zeta+\xi)-2\beta(t)+\xi_0}},\nonumber
\end{equation}
where $\xi_0=-\mathrm{\mathrm{ln}}c$,\ \ \
$r=(x+2ky-12k^2t-\frac{1}{k+q})(x-2qy-12q^2t-\frac{1}{k+q})+\frac{1}{(k+q)^2}$.
\\
{\bf{Example 5}}: (Nondegenerate case) N-soliton solution.

Take the case of $\alpha=i$ for example. We take $u=0$ as the
initial solution of (\ref{d7}) with $m=0$ and let
$$\psi^+_j=e^{k_jx+ik^2_jy-4k^3_jt},\ \ \  \psi^-_{N+j}=\overline{\psi^+_j},\ \ \ k_j\in\mathbb{C},$$
$$C_j(t)=\frac{1}{2Re(k_j)}e^{2\beta_j(t)},\ \ \  j=1,\cdots,N,$$ where
$\beta_j(t)$ are arbitrary functions in $t$. Then the formulas
(\ref{f22}), (\ref{f25}) and (\ref{f26}) give the $N$-soliton
solution of the KPIESCS (\ref{d7}) with degree $N$.

\section*{ Acknowledgment}\hskip\parindent
This work was supported by the Chinese Basic Research Project
"Nonlinear Science".

\hskip\parindent


\begin{thebibliography}{s99}
\bibitem{Mel'nikov88}
Mel'nikov V K 1988 Phys. Lett. A 133 493

\bibitem{Mel'nikov89(1)}
Mel'nikov V K 1989 Commun. Math. Phys. 120 451

\bibitem{Mel'nikov90}
Mel'nikov V K 1990 J. Math. Phys. 31 1106

\bibitem{Leon90(1)}
Leon J, Latifi A 1990 J. Phys. A 23 1385-1403

\bibitem{Leon90(2)}
Leon J 1990 Phys. Lett. A 144 444-452

\bibitem{Doktorov91}
Doktorov E V, Vlasov R A 1991 Opt. Acta 30 3321

\bibitem{Mel'nikov92}
Mel'nikov V K 1992 Inverse Probl. 8 133

\bibitem{Zeng2000}
Zeng Y B , Ma W X and Lin R L 2000 J. Math. Phys. 41 (8) 5453-5489

\bibitem{LinZengMa2001}
Lin R L, Zeng Y B and Ma W X 2001 Physica A  291 287-298

\bibitem{Yeshuo2002}
Ye S, Zeng Y B 2002 J. Phys. A 35 L283-L291

\bibitem{Zeng2001}
Zeng Y B, Ma W X and Shao Y J 2001 J. Math. Phys. 42(5) 2113-2128

\bibitem{Zeng2002}
Zeng Y B, Shao Y J and Ma W X 2002 Commun. Theor. Phys. 38
641-648

\bibitem{Zeng2003}
Zeng Y B, Shao Y J and Xue W M 2003 J. Phys. A  36 1-9

\bibitem{Mel'nikov87}
Mel'nikov V K 1987 Commun. Math. Phys. 112 639-652

\bibitem{Mel'nikov89(2)}
Mel'nikov V K 1989 Commun. Math. Phys. 126 201-215

\bibitem{Zhangdajun2003}
Deng S F, Chen D Y and Zhang D J 2003 J. Phys. Soc. Jap. 72
2184-2192

\bibitem{Ablowitz91}
Ablowitz M J, Clarkson P A 1991 Solitons, Nonlinear Evolution
Equations and Inverse Scattering (Cambridge)

\bibitem{Dickey91}
Dickey L A 1991 Soliton equation and Hamiltonian systems
(Singapore:World Scientific)

\bibitem{Matveev91}
Matveev V B, Salle M A 1991 Darboux Transformations and Solitons
(Berlin: Springer)


\bibitem{Date83}
Date E, Jimbo M, Kashiwara M and Miwa T 1983 In Nonlinear
Integrable Systems-Classical Theory and Quantum Theory, Jimbo M,
Miwa T 1983 (eds.) (Singapore:World Scientific)


\bibitem{Ohta1988}
Ohta Y, Satsuma J, Takahashi D and Tokihiro T 1988 Prog. Theor.
Phys. Suppl. 94 210

\bibitem{Jurij91}
Jurij Sidorenko, Walter Strampp 1991 Inverse Probl. 7 L37-L43

\bibitem{Oevel93}
Oevel W, Strampp W 1993 Commun. Math. Phys. 157 51-81

\bibitem{Dickey95}
Dickey L A 1995 Lett. Math. Phys. 34 379-384

\bibitem{Cheng92}
Cheng Y 1992 J. Math. Phys. 33 3774

\bibitem{Cheng95}
Cheng Y 1995 Commun. Math. Phys. 171 661-682

\end {thebibliography}
\end{document}